\documentclass[letterpaper,twocolumn,showpacs,footinbib,bibnotes,pra,floatfix]{revtex4}
\usepackage{graphicx}
\usepackage{amsmath}
\usepackage{amssymb}
\usepackage{dcolumn}
\usepackage{bm}

\begin{document}
\bibliographystyle{apsrev}

\title{Precision measurement of light shifts at two off-resonant wavelengths in a single trapped Ba$^+$ ion and the determination of atomic dipole matrix elements}
\author{J. A. Sherman}\thanks{Now at Clarendon Laboratory, University of Oxford}\email{jeff.sherman@gmail.com}
\author{A. Andalkar}
\author{W. Nagourney}
\author{E. N. Fortson}
\affiliation{Department of Physics, Box 351560, University of
Washington, Seattle, WA 98195-1560}
\date{\today}

\newcommand{\Sstate}{\ensuremath{6S_{1/2}} }
\newcommand{\Pstate}{\ensuremath{6P_{1/2}} }
\newcommand{\Pstatespecial}{\ensuremath{6P_{1/2,3/2}} }
\newcommand{\Pupper}{\ensuremath{6P_{3/2}} }
\newcommand{\Dstate}{\ensuremath{5D_{3/2}} }
\newcommand{\Dshelve}{\ensuremath{5D_{5/2}} }
\newcommand{\Fstate}{\ensuremath{4F_{5/2}} }
\newcommand{\Ba}{\ensuremath{^{138}\text{Ba}^+} }
\newcommand{\Baodd}{\ensuremath{^{137}\text{Ba}^+} }
\newcommand{\bhat}[1]{\hat{\boldsymbol{#1}}}

\newcommand{\jSectionHeading}[1]{}

\newcommand{\jfrac}[2]{#1/#2}

\begin{abstract}
We have measured the ratio $R$ of the vector ac-Stark effect (or light shift) in the $6S_{1/2}$ and $5D_{3/2}$ states of a single trapped barium ion to 0.2~\% accuracy at two different off-resonant wavelengths.  We earlier found $R = \Delta_S / \Delta_D = -11.494(13)$ at 514.531~nm where $\Delta_{S,D}$ are the vector light shifts of the $6S_{1/2}$ and $5D_{3/2}$ $m = \pm 1/2$ splittings due to circularly polarized light, and now we report the value at 1111.68~nm, $R =  +0.4176(8)$.  These observations together yield a value of the $\langle 5D || er || 4F \rangle$ matrix element, previously unknown in the literature.  Also, comparison of our results with an {\it ab initio} calculation of dynamic polarizability would yield a new test of atomic theory and improve the understanding of atomic structure needed to interpret  a proposed atomic parity violation experiment.  
\end{abstract}

\pacs {32.70.Cs, 32.60.+i, 32.80.Pj}

\maketitle

\begin{figure}
\includegraphics[width=\linewidth]{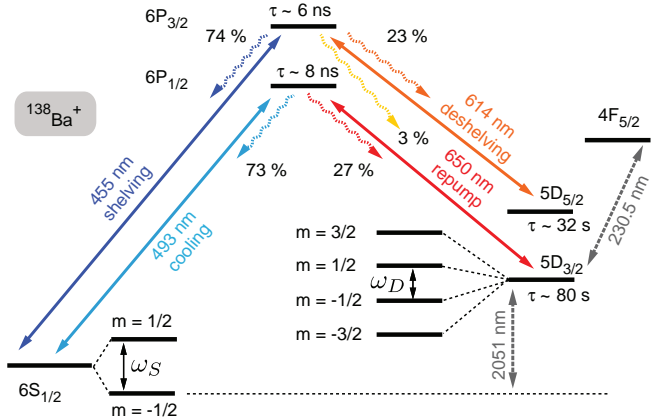}
\caption{This level diagram of Ba$^+$ shows several relevant atomic states, transition wavelengths, lifetimes, and approximate decay branching ratios.  The Zeeman structure of the $6S_{1/2}$ and $5D_{3/2}$ states is shown explicitly;  an external magnetic field makes $\omega_S$ and $\omega_D$ a few MHz.}
\label{fig:energyLevel}
\end{figure}

\section{Introduction}
\begin{table*}
\centering
\caption{Collection of calculated and measured $5D_{3/2}$ and $6S_{1/2}$ dipole matrix elements.  The transition energies are derived from the collection~\cite{curry2004cwe}.  Data from \cite{dzuba2001cpn} are derived from published radial integrals.  No signs are shown for experimental data which are sensitive only to the matrix element magnitude.  Data from~\cite{kastberg1993mat} (see also~\cite{kastberg1994mat}) are derived from absolute transition rates;  data from~\cite{gallagher1967osc,davidson1991osb} are derived from oscillator strengths.  Data presented from~\cite{klose2002cea} assumed LS coupling and rely on various measurements.  A given set of matrix elements generates estimates of the light shift ratio $R$ (see Eq.~\ref{eq:lightshiftRatioEstimate}) to be compared with our measured results $R_\text{514.531~nm} = -11.494(13)$ and  $R_\text{1111.68~nm} = +0.4176(8)$.  Arbitrarily, the estimates of $R$ below use results from~\cite{gopakumar2002edq} when otherwise not available in a given column.  Note the lack of published experimental data on $\langle 5D || er || nF \rangle$ matrix elements.}
\begin{tabular}{cl | ddd | ddd}
					&					& \multicolumn{6}{c}{Dipole matrix elements (units of $e a_0$)} \\
					&					& \multicolumn{3}{c}{Theory} & \multicolumn{3}{c}{Experiment} \\
Transition				& $\Delta E_{ij}$ (cm$^{-1}$)  	& \multicolumn{1}{c}{Ref.~\cite{gopakumar2002edq}}  &   \multicolumn{1}{c}{Ref.~\cite{dzuba2001cpn}} &   \multicolumn{1}{c}{Ref.~\cite{guet1991rmb}}  &  \multicolumn{1}{|c}{Ref.~\cite{kastberg1993mat}} &  \multicolumn{1}{c}{Ref.~\cite{gallagher1967osc}}  & \multicolumn{1}{c}{Ref.~\cite{davidson1991osb,klose2002cea}} \\ \hline \hline
$6S_{1/2}$--$6P_{1/2}$	& 20261.561			& 3.3266            &  3.310        & 3.300        & 3.36(16) 	& 3.36(12)	& 3.36(4)~\mbox{\cite{davidson1991osb}}  \\
$6S_{1/2}$--$6P_{3/2}$	& 21952.404			& -4.6982           & -4.674       & (-)4.658     & 4.45(19)	& 4.69(16)	& 4.55(10)~\mbox{\cite{davidson1991osb}} \\
$6S_{1/2}$--$7P_{1/2}$	& 49389.822			& 0.1193            & -0.099       & & & 							& 0.24(3)~\mbox{\cite{klose2002cea}} \\
$6S_{1/2}$--$7P_{3/2}$	& 50011.340			& -0.3610           & -0.035       & & & 							& 0.33(4)~\mbox{\cite{klose2002cea}} \\
$6S_{1/2}$--$8P_{1/2}$	& 61339.5				& -0.4696	          & -0.115       & & & 						& 0.10(1)~\mbox{\cite{klose2002cea}} \\
$6S_{1/2}$--$8P_{3/2}$	& 61642.0				& 0.5710             & 0.073        & & & 						& 0.15(2)~\mbox{\cite{klose2002cea}} \\ \hline
$5D_{3/2}$--$6P_{1/2}$	& 15387.708			& -2.9449            & 3.055       & 3.009        & 3.03(9)  & 2.99(18)		& 2.90(9)~\mbox{\cite{davidson1991osb}}  \\
$5D_{3/2}$--$6P_{3/2}$	& 17078.552			& -1.2836            & -1.334      & (-)1.312    & 1.36(4) & 1.38(9)		& 1.54(19)~\mbox{\cite{davidson1991osb}} \\
$5D_{3/2}$--$7P_{1/2}$	& 44515.970			& -0.3050            & 0.261       & & &  							& 0.42(11)~\mbox{\cite{klose2002cea}} \\
$5D_{3/2}$--$7P_{3/2}$	& 45137.488			& -0.1645            & -1.472      & & & 							& 0.19(5)~\mbox{\cite{klose2002cea}} \\
$5D_{3/2}$--$8P_{1/2}$	& 56465.6				& -0.1121            & 0.119       & & & 						& 0.23(6)~\mbox{\cite{klose2002cea}} \\
$5D_{3/2}$--$8P_{3/2}$	& 56768.1				& -0.0650            & -0.070      & & & 						& 0.10(3)~\mbox{\cite{klose2002cea}} \\ \hline
$5D_{3/2}$--$4F_{5/2}$	& 43384.765			& -3.69~\mbox{\cite{das2004pc}}	            &                 & & & & \\
$5D_{3/2}$--$5F_{5/2}$	& 52517.070			& 1.59~\mbox{\cite{das2004pc}}                 &                &  & & &\\
$5D_{3/2}$--$6F_{5/2}$	& 59722.48 			& 0.44~\mbox{\cite{das2004pc}}                  &                 & & & & \\ \hline \hline
\multicolumn{2}{l|}{$R_{\lambda = 514.531 \text{ nm}}$ prediction} & \multicolumn{1}{c}{-13.41} & \multicolumn{1}{c}{-12.55} & \multicolumn{1}{c}{-12.75}  & \multicolumn{1}{c}{-13.21} & \multicolumn{1}{c}{-13.35} & \multicolumn{1}{c}{-13.98} \\
\multicolumn{2}{l|}{$R_{\lambda = 1111.68 \text{ nm}}$ prediction} & \multicolumn{1}{c}{0.4444} & \multicolumn{1}{c}{0.4146} & \multicolumn{1}{c}{0.4168} &  \multicolumn{1}{c}{0.5001} & \multicolumn{1}{c}{0.5110} & \multicolumn{1}{c}{0.7444}
\end{tabular}
\label{tab:dipoleMatrixElementTable}
\end{table*}

Although the atomic structure of  singly-ionized barium is generally well understood, and a number of precise calculations of transition matrix elements have been carried out in recent years~(\cite{gopakumar2002edq,saha1993aic,guet1991rmb,safronova:2008tsl}, among others), precise measurements of these matrix elements are relatively few.  Among the low-lying electric dipole transitions, only the $\langle 6S || er || 6P \rangle$ matrix element has been measured with an accuracy approaching 1\%, while the  $\langle 5D || er || 4F \rangle$ matrix element is unknown in the literature (see Table~\ref{tab:dipoleMatrixElementTable}).   Precise measurements are needed to test modern many-body atomic theory in this alkali-like system to the 1\% level and beyond.  Such precision is necessary, for example, to interpret proposed measurements of parity violation in Ba$^+$ and Ra$^+$~\cite{fortson1993pmp,koerber2002rss,koerber2003rfs}.

Earlier we reported~\cite{koerber2002rss,sherman2005pml} developing a technique for evaluating matrix elements by precisely measuring off-resonant Zeeman-like light shifts in a single trapped barium ion.  Here we describe the technique in more detail, and present new measurements that, together with previous results~\cite{sherman2005pml}, provide a more complete picture of matrix elements involving low-lying states of Ba$^+$.  The technique is in principle generalizable to other atoms and ions possessing convenient metastable states; we show that unknown matrix elements can be individually targeted by choosing an appropriate off-resonant light shift wavelength. The key idea is to determine the ratio of light shifts simultaneously
measured in two atomic states, thereby eliminating the need to precisely know or control the light shift laser intensity.  Our measurements place constraints on the Ba$^+$ matrix elements that can test modern atomic theories, such as the coupled-cluster method, and \emph{ab initio} techniques (\cite{gopakumar2002edq,saha1993aic,safronova:2008tsl}, for example).

Through dynamic polarization of a (two-level) atom~\cite{budker2004ape}, an off-resonant light beam shifts the energies of atomic states by
\begin{equation} \label{eq:lightShiftSimpleForm}
\Delta E_{1,2} = \pm \hbar \frac{\Omega^2}{4 \delta},
\end{equation}
where $\Omega$ is the Rabi frequency that scales linearly with the light electric field strength and the atomic dipole matrix element $\langle 1 | er | 2 \rangle$, and $\delta$ is the detuning of the light beam from the two state resonance frequency.  Eq.~\ref{eq:lightShiftSimpleForm} is easily generalized to multi-state atoms.  The light shift of a state $|\gamma, j, m\rangle$ due to an oscillating field $\boldsymbol{E}(t) = \boldsymbol{E} \cos \omega t$ is, to second-order in perturbation theory \cite{stalnaker2006dse},
\begin{equation} \label{eq:lightShiftSimpleFormMatrixElements}
\begin{split}
\Delta E_{\gamma, j, m} &= \frac{(E)_i(E)_k}{4} \sum_{\gamma', j', m', \pm} \langle \gamma,j,m | e r_i | \gamma', j', m' \rangle \\
 & \times \langle \gamma',j',m' | e r_k | \gamma, j, m \rangle \\ 
 & \times \frac{1}{E_{\gamma,j,m} - E_{\gamma',j',m'} \pm \hbar \omega},
\end{split}
\end{equation}
where $\boldsymbol{E}$ is the electric field strength times a polarization vector $\boldsymbol{\epsilon}$, $e$ is the electron charge, and summation over the indices $i$ and $k$ is implied.  The term containing $+ \hbar \omega$ in the denominator (termed the `Bloch-Siegert' shift in the context of radio frequency spectroscopy) is often ignored when the rotating wave approximation is made~\cite{cohen-tannoudji1992api} but must be included here.

%

The next order in perturbation theory, called \emph{hyperpolarizability}, is unimportant unless the supposedly off-resonant laser is accidentally resonant with a two-photon transition~\cite{brusch2006hes}.  We empirically find no hyperpolarizability effects by confirming that ratios of light shifts do not change with laser intensity.

\begin{figure}
\centering
\includegraphics[width=\linewidth]{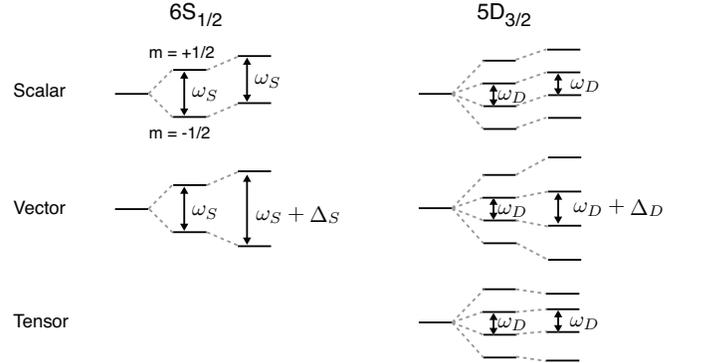}
\caption[Structure of the multipole light shifts in the $6S_{1/2}$ and $5D_{3/2}$ states.]{Structure of the multipole light shifts in the $6S_{1/2}$ and $5D_{3/2}$ states.  In this schematic, we imagine a magnetic field splitting the $6S_{1/2}$ and $5D_{3/2}$ sublevels by $\omega_S$ and $\omega_D$ in the vector structure of the Zeeman interaction.  Further (assumed small) perturbations due to the ac-Stark effect add scalar, vector, and tensor-like shifts.}
\label{fig:lightShiftMultipoleCartoon}
\end{figure}

If the light shifts are much smaller than the Zeeman energies, the resulting energy shift also can be written in terms of tensor ranked polarizabilities  $\alpha_0$, $\alpha_1$, and $\alpha_2$ \cite{stalnaker2006dse}:
\begin{equation}
\begin{split}
\Delta E_{\gamma, j, m} &= - \underbrace{\frac{\alpha_0}{2} | \boldsymbol{E} |^2}_\text{scalar}
- \underbrace{i \frac{\alpha_1}{2} \frac{m}{j} (i |\boldsymbol{E} \times \boldsymbol{E^*}|) }_\text{vector} \\
&- \underbrace{\frac{\alpha_2}{2} \left( \frac{3m^2 - j(j+1)}{j(2j-1)}\right) \frac{3 |E_z|^2 - |\boldsymbol{E}|^2}{2}}_\text{quadrupole/tensor},
\end{split}
\end{equation}
which is identical to the form of the dc-Stark effect except for the addition of a vector-like term.  In general, the vector shift term is maximal for pure circularly polarized light aligned with any existing magnetic field.  In other words, both the relative strength of circular polarization 
\begin{equation}
| \sigma_z | = \left| \frac{\boldsymbol{E} \times \boldsymbol{E^*} \cdot \bhat{z}}{E^2} \right|,
\end{equation}
and the vector light shift term are maximized when $\boldsymbol{E} = E (\bhat{x} \pm i \bhat{y}) / \sqrt{2}$.  Other researchers call this the `Zeeman-like' ac-Stark shift~\cite{park2001mzl} or a vector shift specified by a `fictitious' magnetic field~(\cite{haroche1970mzh}, \footnote{Such light shifts have proven to be practical replacements for a magnetic field:  e.g., an ac-Stark optical Stern-Gerlach effect~\cite{park2002osg}, an optically induced Faraday effect~\cite{cho2005oif}, an optical Feshbach resonance~\cite{dickerscheid2005fro}, an ac-Stark phase shift compensator akin to a spin-echo~\cite{haffner2003pma}, even a technique to enable highly forbidden $J = 0 \leftrightarrow 0$ optical `clock' transitions in optical lattice trapped atoms~\cite{ovsiannikov2007mwi}.}).

\begin{table}
\centering
\caption[Scalar, vector, and tensor coefficients used in the light shift ratio estimate]{Scalar, vector, and tensor coefficients used in estimating the light shift ratio shift magnitudes in the $m = \pm 1/2$ splittings in the $6S_{1/2}$ and $5D_{3/2}$.  In particular, the constants $v(j,j')$ are necessary in the computation of our measured light shift ratios in Eq.~\ref{eq:lightshiftRatioEstimate}.}
\begin{tabular}{c | c  c  c}
	   & \multicolumn{3}{c}{Relevant light shift coefficients} \\
$(j, j')$ & Scalar $s(j,j')$ & Vector $v(j,j')$ & Tensor $t(j,j')$ \\ \hline \hline
(1/2, 1/2) & -1/6 & -1/3 & 0 \\
(1/2, 3/2) & -1/2 & 1/6 & 0 \\
(3/2, 1/2) & -1/12 & -1/12 & 1/12 \\
(3/2, 3/2) & -1/12 & -1/30 & -1/15 \\
(3/2, 5/2) & -1/12 & 1/20 & 1/60
\end{tabular}
\label{tab:lightShiftCoefficients}
\end{table}

\begin{figure}
\centering
\includegraphics[width=\linewidth]{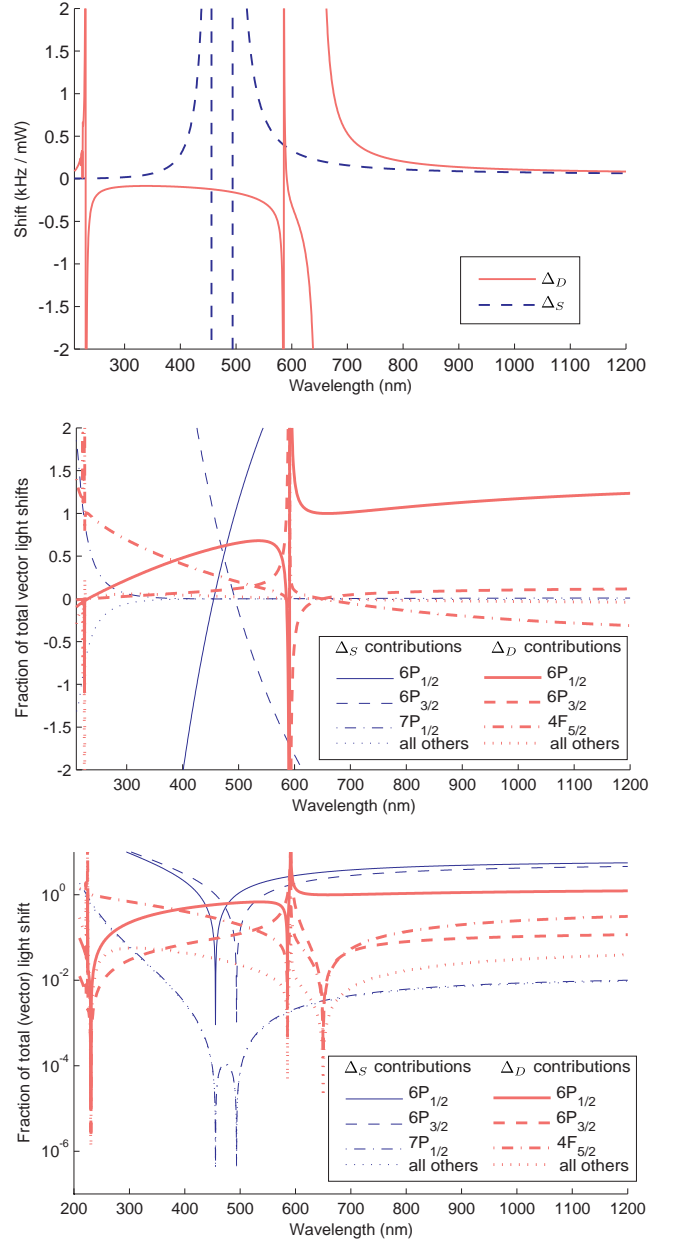}
\caption{(top) Estimated total light shifts $\Delta_S$ and $\Delta_D$ assuming a 20~$\mu$m, circularly polarized laser beam.  (middle) The fractional sizes and signs of contributions to the vector light shifts $\Delta_S$ and $\Delta_D$ due to various nearby states,  shown on a linear and (bottom) a logarithmic scale plots.  Notice, for instance, that in the visible spectrum, the dominant contribution to the $6S_{1/2}$ vector shift $\Delta_S$ is due to the large and oppositely signed contributions of $6P_{1/2}$ and $6P_{3/2}$.}
\label{fig:lightShiftContrib}
\end{figure}

We now focus on the $6S_{1/2}, m= \pm 1/2$ and $5D_{3/2}, m= \pm 1/2$ splittings $\omega_S$ and $\omega_D$ (refer to Figure~\ref{fig:energyLevel}) and their associated light shifts $\Delta_S$ and $\Delta_D$.  As shown in Figure~\ref{fig:lightShiftMultipoleCartoon}, only vector pieces of the ac-Stark effect can accomplish such shifts.  Therefore, scalar and tensor light shifts are not directly measured (though we will see that the tensor piece in the $5D_{3/2}$ state plays a role in small systematic $\omega_D$ resonance line-shape distortions).  If we define the shifted resonances
\begin{align}
\omega_S^{LS} &= \omega_S + \Delta_S, \\
\omega_D^{LS} &= \omega_D + \Delta_D, \label{eq:deltaDdef}
\end{align}
then we can form a quantity we call the light shift ratio from the measurement of two shifted and two unshifted resonance measurements
\begin{equation}
\label{eq:Ratiodef}
R \equiv \frac{\Delta_S}{\Delta_D} = \frac{\omega_S^{LS} - \omega_S}{\omega_D^{LS} - \omega_D}.
\end{equation}
Writing Eq.~\ref{eq:lightShiftSimpleFormMatrixElements} in terms of atomic dipole reduced matrix elements, the light shift ratio is
\begin{align}
R = \frac{\Delta_S}{\Delta_D} &= \frac{\Delta E_{6S_{1/2}, m=1/2} - \Delta E_{6S_{1/2}, m=-1/2}}{\Delta E_{5D_{3/2}, m=1/2} - \Delta E_{5D_{3/2}, m=-1/2}} \\
	&= \frac{\displaystyle \sum_{\gamma', j', \pm} v(\tfrac{1}{2},j') \frac{ | \langle 6S_{1/2} || er || \gamma', j' \rangle |^2}{(E_{\gamma',j'} - E_{6S_{1/2}} \pm \hbar \omega)}}{ \displaystyle \sum_{\gamma', j', \pm} v(\tfrac{3}{2},j')  \frac{| \langle 5D_{3/2} || er || \gamma' j' \rangle |^2}{(E_{\gamma', j'} - E_{5D_{3/2}} \pm \hbar \omega)}}, \label{eq:lightshiftRatioEstimate}
\end{align}
where $E_{\gamma', j'}$ is the energy of state $|\gamma', j' \rangle$, $\omega = 2 \pi / \lambda$ is the off-resonant laser frequency, and $v(j,j')$ is a vector light shift coefficient calculated from Clebsch-Gordan coefficients and tabulated with similar scalar and tensor shift coefficients in Table~\ref{tab:lightShiftCoefficients}.  In Section~\ref{sec:extractionOfMatrixElements} we will use Eq.~\ref{eq:lightshiftRatioEstimate} to interpret our measurements.  One result will be improved values for some of the dipole reduced matrix elements $\langle 6S_{1/2} || er || \gamma', j' \rangle$ and $\langle 5D_{3/2} || er || \gamma', j' \rangle$.  Current theoretical calculations and previous measurements of these are shown in Table~\ref{tab:dipoleMatrixElementTable}, and are sufficiently accurate to compute an estimate of the light shifts $\Delta_S$ and $\Delta_D$ as a function of laser wavelengths.  We plot the results in Figure~\ref{fig:lightShiftContrib}.  Assuming pure circular polarization and a 20~$\mu$m laser spot centered on the ion, observed shifts are approximately $\Delta_S = 2.59$~kHz/mW, $\Delta_D = -0.187$~kHz/mW at $\lambda =$~514.531~nm, and  $\Delta_S = 0.069$~kHz/mW, $\Delta_D = 0.093$~kHz/mW at $\lambda =$~1111.68~nm.

\section{Apparatus}
As described in earlier reports~\cite{koerber2003rfs,sherman2005pml}, we trap single barium ions in a twisted-wire Paul-Straubel ring trap roughly 0.75~mm in diameter.  The trap itself is made from tantalum, and all nearby surfaces---vacuum feedthrough pins, ovens containing pure barium, dc compensation electrodes, an electron emission filament, and a half loop of wire that serves as a source of spin-flip radio frequency fields---are made from non magnetic tungsten or tantalum.  A uniform $\sim 1.5$~G magnetic field, established by coils mounted on the vacuum system, splits magnetic sub-levels by a few MHz and avoids a fluorescence dark state~\cite{berkeland2002dds}.  Surrounding the vacuum chamber and magnetic field coils are two layers of magnetic shielding.  Un-lightshifed spin-resonance resolution is ultimately limited by residual magnetic noise of $\sim 40 \text{ $\mu$G}/\sqrt{\text{Hz}}$ at the site of the ion.

We perform Doppler cooling on the 493~nm $6S_{1/2} \leftrightarrow 6P_{1/2}$ transition while also applying a 650~nm laser to prevent the ion from populating the metastable $5D_{3/2}$ state.  Both of these lasers are locked to an opto-galvanic resonance in a barium hollow cathode lamp; the 493~nm laser linewidth is typically 10~MHz while the 650~nm laser linewidth is purposefully broadened by modulation to $\sim 100$~MHz to destroy a coherent population trapping dark state in this system~\cite{janik1985dfo}.  Camera optics image the ion fluorescence onto a photo-multiplier tube;  we typically observe 4000 counts/s of 493~nm fluorescence when slightly above saturation.  Filtered light emitting diodes (Lumileds) are light sources that cover the 455~nm $6S_{1/2} \leftrightarrow 6P_{3/2}$ transition that can result in a decay to the metastable $5D_{5/2}$ `shelving' state, and the 614~nm $5D_{3/2} \leftrightarrow 6P_{3/2}$ transition that correspondingly empties or `deshelves' the $5D_{5/2}$ state.  Though the diodes' spectra are extremely broad, we nonetheless can drive these transitions at rates exceeding 10~Hz.  (We have driven the $6S_{1/2} \leftrightarrow 5D_{5/2}$ electric-quadrupole transition at 1762~nm, most recently with a diode pumped fiber laser source, which would provide more efficient shelving/deshelving in future experiments.)

\section{Single ion rf spectroscopy}
We have developed a technique to measure the $6S_{1/2}$ and $5D_{3/2}$  $m = \pm 1/2$ spin resonance frequencies $\omega_S$ and $\omega_D$ (refer to Figure~\ref{fig:energyLevel}) in a single trapped ion.  After a period of laser cooling, the 493~nm beam is made dim and circularly polarized via an acousto-optic and electro-optic modulator.  Since the laser beam is aligned with the dominant magnetic field, optical pumping into one of the $6S_{1/2}$ sub-levels results with extremely high efficiency;  for this discussion we assume pumping into the $m = -1/2$ state.  Both the 493~nm and 650~nm resonant lasers are shuttered, and a precisely-timed pulse of radio-frequency voltage is applied to a half-loop of wire about 1~mm away from the trapped ion, oriented such that an oscillating magnetic field perpendicular to the quantization axis develops.  This pulse is engineered to be the appropriate duration and strength to be a so-called $\pi$-pulse;  when resonant, it causes the ion spin to completely flip from $m = -1/2$ to $m= +1/2$.  Then, a dim, circularly polarized `probe' pulse of 493~nm light either leaves the ion  alone if it remained in the $6S_{1/2}$, $m = -1/2$ state after the rf pulse, or moves it to the $6P_{1/2}$ if the rf pulse succeeded in flipping the spin to $m = +1/2$.  Roughly 30 \% of the time, this excited ion will then decay to the long-lived $5D_{3/2}$ state.  A 150~ms pulse of 455~nm light moves any ion still in the $6S_{1/2}$ state to the $5D_{5/2}$ via a decay from $6P_{3/2}$ state (the decay fraction to $5D_{3/2}$ is fortunately very rare, approximately 3\%).  Finally, application of bright, linearly polarized 493~nm and 650~nm radiation results in ion fluorescence only if the ion is not stuck in $5D_{5/2}$;  this is correlated to whether the rf spin flip was successful.

\begin{figure}
\centering
\includegraphics[width=\linewidth]{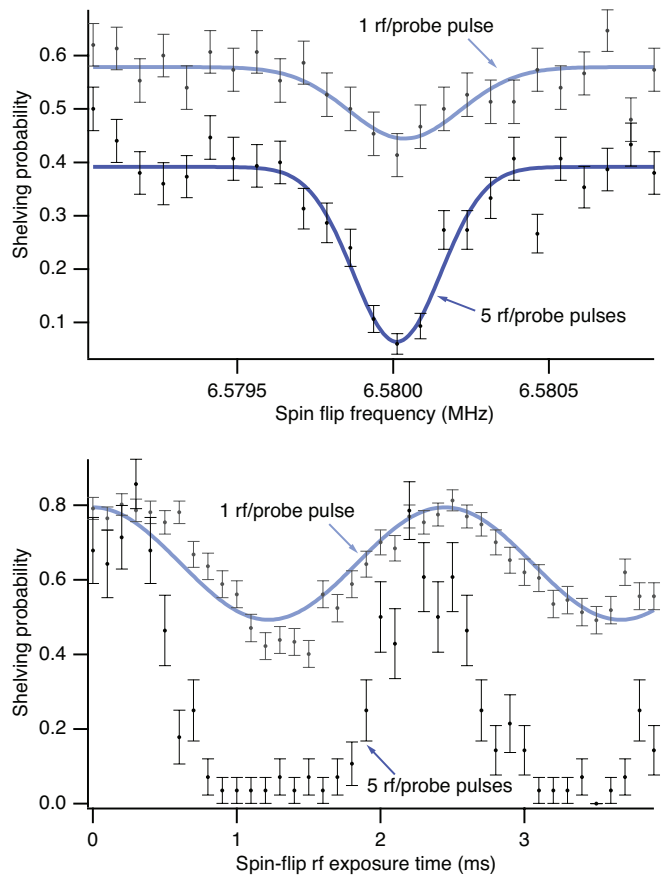}
\caption{Single ion spin resonances in the $6S_{1/2}$ state as functions of frequency (top) and exposure time (bottom).  See text for a description of the spin sensitive electron shelving technique.  We found that the low contrast of the spin resonances (due to an unfavorable branching ratio of decays from $6P_{1/2}$) could be improved by exposing the ion to several repeated rf spin flips and optical probe pulses as shown here.}
\label{fig:sResPulsesCompare}
\end{figure}
This measurement sequence is repeated many times to gather statistics, and using several trial spin flip radio frequencies to build a resonance curve as shown in Figure~\ref{fig:sResPulsesCompare}; varying the rf pulse duration while on resonance shows the Rabi probability flopping.  The contrast in such curves is low due to the 30\% branching ratio we rely on to transfer spin information of the the $6S_{1/2}$ into state information in the $5D_{3/2}$ state.  We found that we could improve the contrast of the spin resonances by applying repeated rf spin slip and dim 493~nm optical probe pulses at the expense of the coherent features of the resonance curves.  We found that 3--5 repetitions yielded the most success;  two repetitions mimicked certain aspects of Ramsey interrogation, complicating curve fitting, while many more repetitions led to broadening of resonances.  We aggressively searched for, and did not find at the level of $10^{-3}$ of the light shift, any shifts in the resonance centers due to the additional pulses.

\begin{figure}
\centering
\includegraphics[width=\linewidth]{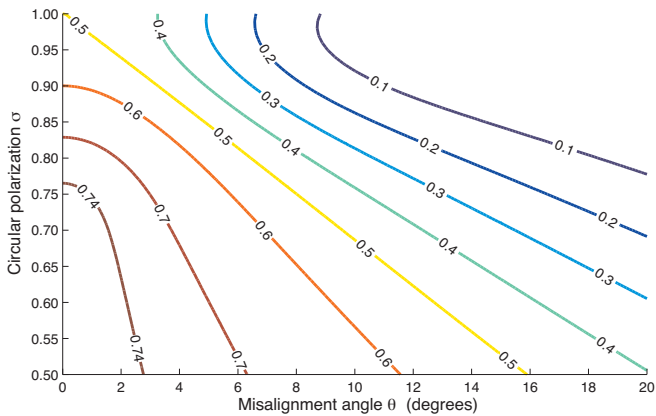}
\caption{A rate equation model of optical pumping allows us to estimate the steady state population $a_{-1/2}$ of the $5D_{3/2}, m = -1/2$ state when linearly polarized 493~nm and $\sigma^{-}$ polarized 650~nm light is incident on the single ion.  We plot contours of $a_{-1/2}$, an important parameter for a systematic lineshape effect, against the misalignment angle of the 650~nm beam to the magnetic field $\theta$, and the strength of circular polarization $|\sigma|$.  Notice that $a_{-1/2}$ is maximized with imperfect polarization.}
\label{fig:opticalPumpingD}
\end{figure}

\begin{figure}
\centering
\includegraphics[width=\linewidth]{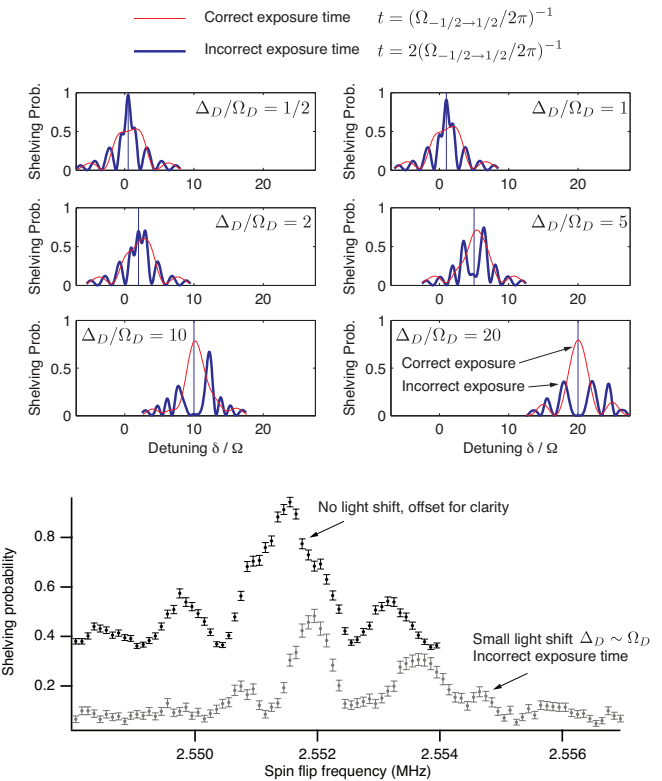}
\caption[Distortion in light shift data from incorrect exposure timing]{These data demonstrate theoretically (top) and experimentally (bottom) distortions in light shifted $5D_{3/2}$ spin resonance curves resulting from using the same spin-flip exposure time for large and small light shifted resonances.  At large light shifts $\Delta_D \gg \Omega_D$, the $5D_{3/2}, m= \pm 1/2$ levels are isolated into an effective two level system with a different Rabi frequency than the unshifted $J = 3/2$ system~\cite{sherman2007thesis,shore1990tca2}.  At small light shifts $\Delta \sim \Omega_D$, the tensor light shift makes all the $5D_{3/2}$ splittings quasi resonant, leading to complicated lineshapes.}
\label{fig:distortedLightShiftExample}
\end{figure}

We employ a similar technique to measure the $5D_{3/2}, m = \pm 1/2$ splitting $\omega_D$.  After a period of laser cooling, the 650~nm laser beam is made circularly polarized; for instance, $\sigma^-$.  Population accumulates through optical pumping in the $5D_{3/2}$ $m=-1/2$ and $m=-3/2$ levels.   The fraction of population in $m=-1/2$, which we call $a_{-1/2}$, turns out to be an important parameter of systematic effects discussed later.  Figure~\ref{fig:opticalPumpingD} shows the result of a rate equation model for determining $a_{-1/2}$ for a given 650~nm laser misalignment and degree of circular polarization;  notice that $a_{-1/2}$ is maximized with imperfect polarization.  With the 650~nm and 493~nm lasers shuttered, a timed radio frequency pulse is engineered to transfer population from the $m = -1/2, -3/2$ manifold to the $m= +1/2,+3/2$ manifold of $5D_{3/2}$.  Though the four-state spin flip dynamics are different from the well known $J=1/2$ case, an effective `$\pi$-pulse' can nonetheless be defined that results in maximum population transfer in analogy to a spin-flip~\cite{shore1990tca2}.  The dim, circularly polarized 650~nm laser again probes the $5D_{3/2}$ state, transferring any population in the upper spin manifold out to the ground state via a likely decay from $6P_{1/2}$ while an ion left in the $m=-1/2$ or $m=-3/2$ state is likely unaffected.  A 455~nm shelving pulse moves an ion in $6S_{1/2}$ to the metastable $5D_{5/2}$ state, as before.  Finally, resonant 650~nm and 493~nm laser light results in bright ion fluorescence as long as the ion is not in the $5D_{5/2}$ shelved state; dim observed fluorescence is correlated with the success of the rf spin flip.

As seen in previous work~\cite{sherman2005pml}, and Figure~\ref{fig:distortedLightShiftExample}, the shape of the curves differs from the familiar $\sin^2 \Omega t$ dependence relevant in the $J = 1/2$ case, but can be fit to the solution of the appropriate optical Bloch equations~\cite{koerber2003thesis}.  Some details of the resonance curves depend sensitively on the initial condition $a_{-1/2}$ of the $5D_{3/2}$ state, as well as the duration, polarization, and alignment of the 650~nm probe pulse.

\section{Measurement of light shifts}
\begin{figure}
\centering
\includegraphics[width=\linewidth]{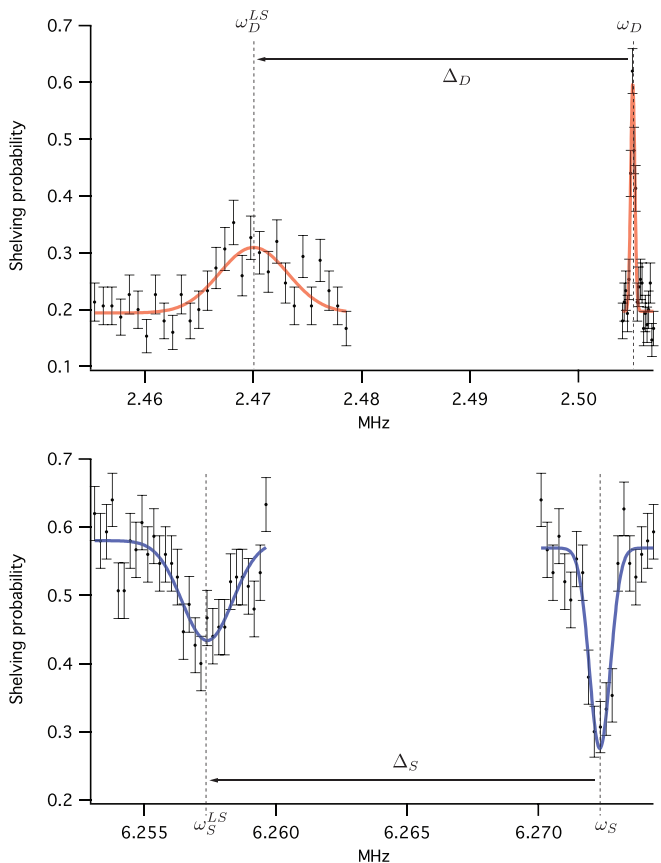}
\caption{In this example spin resonance data set, we find the Zeeman splittings $\omega_S$ and $\omega_D$ with and without application of an off-resonance light shift laser.  Though coherent features are observed on the spectral lines, in practice we fit the resonance profiles to Gaussian lineshapes since temporal magnetic drift and light shift laser intensity, pointing, and polarization noise tends to broaden the lines.}
\label{fig:fourResExample}
\end{figure}
Here we describe our measurements of the vector light shifts $\Delta_S$ and $\Delta_D$ due to an applied beam of off-resonant, circularly polarized light.  The ratio $R = \Delta_S / \Delta_D$ (see Eq.~\ref{eq:Ratiodef}--\ref{eq:lightshiftRatioEstimate}) is independent of the intensity of the laser, and largely independent of errors in polarization and alignment;  see Figure~\ref{fig:fourResExample} for an example data run.   A previous report~\cite{sherman2005pml} details our result using a single-frequency Argon-ion laser at 514.531~nm.  We will now discuss a new measurement using a 1W, polarization-preserving, Yb-fiber laser (Koheras) at 1111.68~nm.  Though tunable, we operated this laser with no active frequency stabilization;  its wavelength was checked with a commerical wavemeter between measurements so that laser frequency variations would cause only a $10^{-5}$ relative error in our reported value of $R$.  After warm-up, the laser's long term power variation, measured after a high quality polarizer, was less than 2\%.  The light shift beam polarization was made circular with an achromatic quarter-wave plate ($|\sigma| > 0.9$) or Babinet-Solieil compensator ($|\sigma| > 0.95$).  The beam is aligned to counter-propagate along the 493~nm and 650~nm resonant beams, and therefore also along the magnetic field, confirmed to within $5^\circ$ by separate optical pumping experiments.

We took interleaved resonance scans of $\omega_S$, $\omega_D$, $\omega_S^{LS}$ and $\omega_D^{LS}$, the latter being spin-resonance scans with the light shift laser incident on the ion.  The scans $\omega_S^{LS}$ and $\omega_D^{LS}$ differed from $\omega_S$ and $\omega_D$ in only two respects.  First, because the residual power, pointing, and polarization noise of the light shift laser causes drift in the frequencies $\omega_S^{LS}$, and $\omega_D^{LS}$, we used higher spin-flip radio frequency voltages during the light shifted scans to power-broaden the resonances by factors of 2--10;  however, the radio frequency pulse `area' $\Omega_S  t$ was kept constant.  Second, the tensor part of the light shift in the $5D_{3/2}$ state made the $5D_{3/2}, m= \pm 1/2$ levels an effective two-level system for light shifts large compared to the spin flip Rabi frequency; i.e.\ $\Delta_D \gg \Omega_D$.  In this limit, the spin-dynamics of the $J = 3/2$ system changes such that the effective $\pi$-pulse exposure time scales down by a factor of 2 compared to the unshifted $5D_{3/2}$ sublevels~\cite{sherman2007thesis};  see Figure~\ref{fig:distortedLightShiftExample}. Therefore, the corect spin-flip rf pulse `area' $\Omega_D^{LS} t_D^{LS} = \Omega_D t_D /2$.

\section{Data}
\begin{figure}
\centering
\includegraphics[width=\linewidth]{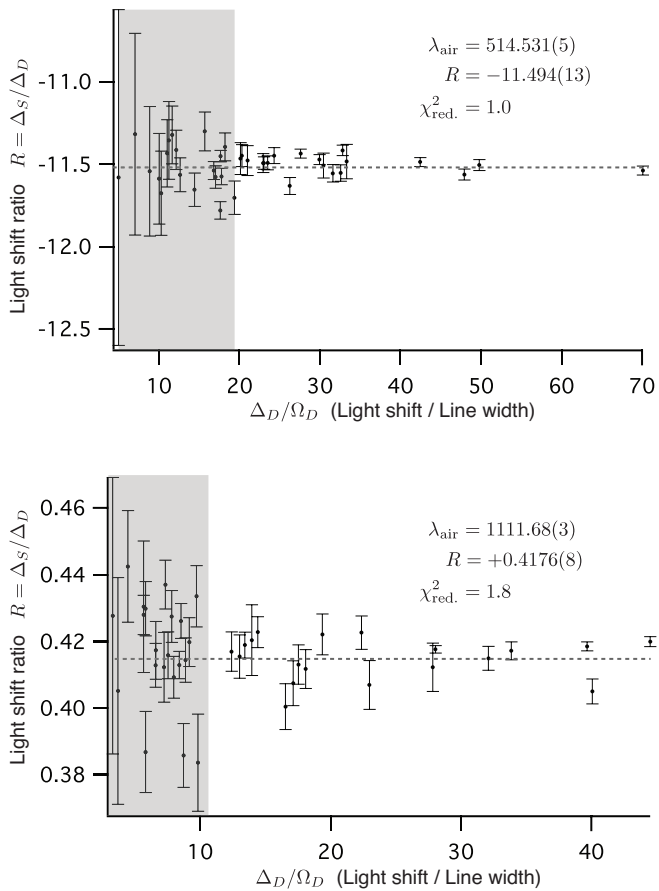}
\caption{Light shift ratio data at 514~nm and 1111~nm plotted against the most important parameter for systematic effects, $\Delta_D/\Omega_D$ (the light shift magnitude over the $5D_{3/2}$ resonance linewidth).  Data in the shaded regions are cut from the straight line fit because the size of the lineshape fitting error is expected to be $> 10^{-3}$ while the sign is uncontrolled;  this explains the large scatter at low $\Delta_D/\Omega_D$.  In each case, this data cut does not change the fitted values for $R$ beyond the quoted errors. No statistically significant slopes are derived from these plots which might otherwise imply an effect due to hyperpolarizability.  To account for the larger than expected scatter in the 1111~nm data, we have $\chi^2$-corrected the reported value.}
\label{fig:lightShiftRatioData}
\end{figure}
A typical resonance curve scan consisted of a few hundred spin-flip trials at about 20 equally spaced radio frequencies, 10 or so of which would be on or near-resonant.  About 1~hour of data taking would reach $\sim 1$~Hz statistical resolution on the few MHz $\omega_S$ and $\omega_D$.  Four to eight hours of data taking would yield $\sim 0.1\%$ statistical precision on the quantity $R$;  one such data collection run is shown in Figure~\ref{fig:fourResExample}.  Figure~\ref{fig:lightShiftRatioData} shows the results of about 40 of these runs at each wavelength, plotted against $\Delta_D / \Omega_D$, the most important parameter leading to systematic effects.  The fitted slope of $R$ vs.\ $\Delta_D$ is zero within the statistical errors, implying that the light shift ratio is independent of laser intensity.  Adding statistical (one standard deviation) and systematic errors (discussion to follow) together in quadrature, we find that
\begin{align}
R(514.531~\text{nm}) &= -11.494(13), \\
R(1111.68~\text{nm}) &= +0.4176(8).
\end{align}
The quoted error in the latter result has been $\chi^2$-corrected by a factor of 1.3 to account for unexplained excess scatter in the data;  the reduced $\chi^2$ of the result at 514.531~nm is 1.0.

\section{Systematic error analysis}
A leading virtue of the experiment design is the lack of significant systematic effects.  Light shifts are proportional to the laser intensity $I$, which in general is difficult to measure at the site of the ion with any precision.  By measuring the light shifts in two states, however, the intensity drops out completely in the ratio $R$.  By choosing to measure the $m = \pm 1/2$ splittings in both the $6S_{1/2}$ and $5D_{3/2}$, errors in laser polarization and alignment affect the light shift ratio only in second order.  By interleaving the measurements temporally, noise and drift in the magnetic field, laser pointing, polarization, and intensity are shared by both shifted and unshifted resonances and thus do not contribute to the ratio.

The effects that remain are smaller than the statistical precision of the data.  The first effect documented in this section is the most important:  line-shape effects that plague the light-shifted $5D_{3/2}$ resonance.  We then consider laser problems:  misalignments, polarization errors, fluctuations, and spectral purity.  Next, we analyze ac-Zeeman effects from the trapping and spin-flip fields, and finally, systematic fluctuations in the magnetic field and ion position.

\subsection{$5D_{3/2}$ resonance line shape effects} \label{sec:lightShiftLineShapes}

Though we expect to measure the $m = \pm 1/2$ splitting $\Delta_D$ in $5D_{3/2}$, the shifts of the outer two splittings
\begin{align}
\Delta_\text{upper} &\equiv E_{5D_{3/2}, m=3/2} - E_{5D_{3/2}, m=1/2} - \omega_D,\\
\Delta_\text{lower} &\equiv  E_{5D_{3/2}, m=-1/2} - E_{5D_{3/2}, m=-3/2}- \omega_D
\end{align}
nevertheless play a role in shifting the $\omega_D^{LS}$ lineshape through distortions of the resonance profile.  Only the vector part of the light shift Hamiltonian shifts magnetic sublevels an amount proportional to $m$.  While scalar shifts are completely undetectable, the tensor shift makes  $\Delta_\text{upper}$ and  $\Delta_\text{lower}$ in general different than $\Delta_D = \Delta_\text{middle}$.  If the light shifts are comparable in size to the spin-flip Rabi frequency $\Delta_D \sim \Omega$, then all the shifted $5D_{3/2}, m \leftrightarrow m \pm 1$ resonances are quasi-resonant with the spin-flip field.  Since our optical pumping method prepares the $5D_{3/2}$ state in some statistical mixture of the $m = -1/2, -3/2$ states, the resulting spin resonance curves can be quite complicated and distorted as illustrated in Figure~\ref{fig:distortedLightShiftExample}.  We expect the size of the distortion to decrease as the relative size of the vector light shift $\Delta_D / \Omega_D$ increases;  this is confirmed with a model of the Bloch equations.  Also, the \emph{sign} of the distortion reverses if the sense of circular polarization of either the 650~nm optical pumping beam, or the light shift beam polarization reverses;  this explains the large scatter of the data in Figure~\ref{fig:lightShiftRatioData} at small relative light shifts $\Delta_D/\Omega_D$.  For data collected with the 1111.68~nm laser, we performed an equal number of data runs in each of the four possible light shift polarization and $5D_{3/2}$ state preparation configurations.

\begin{figure}
\centering
\includegraphics[width=\linewidth]{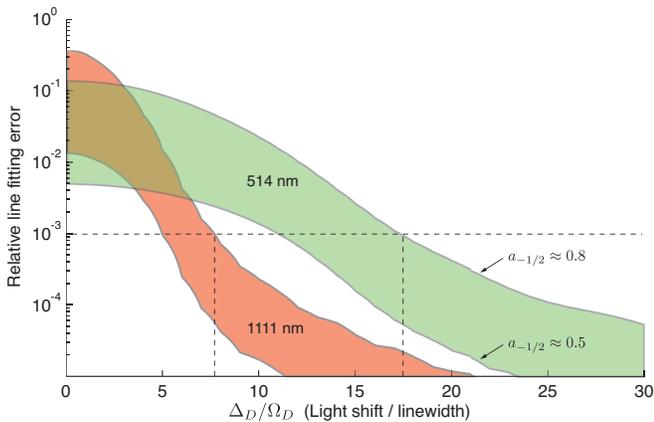}
\caption[Models of the $5D_{3/2}$ spin resonance line-fitting error]{Models of the $5D_{3/2}$ spin resonance line-fitting error.  The top graph compares the typical error magnitude for 514~nm and 1111~nm light shifts considered here.  The bottom graph shows the error profiles for some other candidate light shift wavelengths.  For each wavelength, the top curve assumes initial $5D_{3/2}$ optical pumping fraction $a_{-1/2} = 0.8$, the maximum typically observed while the bottom curve assumes $a_{-1/2} = 0.51$ which minimizes lineshape error.  Dashes lines illustrate approximate bounds on $\Omega_D / \Delta_D$ that keep the line fitting relative error below 0.1\%.}
\label{fig:linefittingModel}
\end{figure}

We observe that the \emph{size} of the distortion depends sensitively on the optical pumping conditions that prepare the ion, while the \emph{sign} depends on the sense of light shift circular polarization $\sigma < 0$ or $\sigma >0$, and whether the ion is initially prepared in the $m = -1/2,-3/2$ or $m = +1/2,+3/2$ manifolds by choosing a $\sigma^-$ or $\sigma^+$ polarized 650~nm beam during optical pumping.  We modeled the optical pumping, light shift Hamiltonian, rf spin-flip, and detection-by-shelving aspects of the experiment by numerically integrating the Bloch equations with the spin flip Hamiltonian using Runge-Kutta routines~\cite{press1992nrc} for all conceivable experimental conditions at several candidate light shift wavelengths.  Simulated experimental data were then subjected to our curve fitting routines to determine $\Delta_D$ as we would measure it.  The expected relative $\Delta_D$ shift error is plotted in Figure~\ref{fig:linefittingModel} against the relevant scaling parameter:  the ratio of light shift to spin-flip Rabi frequency $\Delta_D / \Omega_D$.  Intuitively, for $\pi$-pulses of rf, $\Delta_D / \Omega_D$ is also a measure of the vector light shift divided by the spin-flip linewidth.

Since the sign of the error depends on both the sign of the light shift laser polarization and the sign of the circular polarization used for state preparation, in principle the error can be cancelled by performing reversals of these parameters.  While we did implement such reversals in the 1111~nm data set, the earlier runs using 514~nm were not systematically controlled in this way.  Therefore, for a consistent treatment, we have decided to cut data which our model suggests has a line fitting error of more than $0.1~\%$.  In both data sets, this cut procedure does not end up shifting the reported $R$ values by more than the stated final precision.

\subsection{Laser misalignments, polarization errors}
Our experiment is designed to ideally measure only \emph{vector} light shifts which are maximized in the case of pure circular polarization aligned along the magnetic field.  We will show in this section that the light shift ratio $R$ is insensitive to misalignments and errors in polarization as long as the light shifts are much smaller than the Zeeman shifts,
\begin{align*}
\Delta_S &\ll \omega_S, \\
\Delta_D &\ll \omega_D.
\end{align*}

First we will treat laser misalignments and polarization errors following~\cite{koerber2003thesis,schacht2000thesis}.  If a (positive) circularly polarized laser beam along $\boldsymbol{k}$ is misaligned by a small angle $\alpha$ with respect to the magnetic field $\boldsymbol{B}$, the polarization in a spherical basis becomes
\begin{equation*}
\boldsymbol{\epsilon} = \frac{1}{\sqrt{2}} \begin{pmatrix} \cos \alpha \\ 0 \\ \sin \alpha \end{pmatrix}.
\end{equation*}
If we further allow a small relative phase $\delta$ between the orthogonal components $\epsilon_x$ and $\epsilon_y$, the light becomes elliptically polarized,
\begin{equation*}
\boldsymbol{\epsilon} = \frac{1}{\sqrt{2}} \begin{pmatrix} \cos \alpha \\ ie^{i \delta} \\ \sin \alpha \end{pmatrix}.
\end{equation*}
The size of the vector part of the light shift is proportional to the degree of circular polarization strength defined by
\begin{align}
\sigma &= | \boldsymbol{\sigma} | = | \boldsymbol{\epsilon}^* \times \boldsymbol{\epsilon} |, \\
& = |(-\cos \delta \sin \alpha, 0, \cos \delta \cos \alpha) |.
\end{align}
Some refer to the vector part of the light shift as an effective magnetic field along the vector $\boldsymbol{\sigma}$.  When $\alpha = 0$, $\boldsymbol{\sigma}$ points along the direction of the main magnetic field $\bhat{z}$, and the quantity $\Delta_D$ defined in Eq.~\ref{eq:deltaDdef} exactly measures the vector light shift.  But for $\alpha \ne 0$, the light shift perturbation does not commute with the Zeeman Hamiltonian.  Specifically, the total Hamiltonian in the $5D_{3/2}$ state is
\begin{widetext}
\begin{align}
H &= H_\text{Zeeman} + H_\text{Light shift} \\
&= \scriptsize \begin{pmatrix}
-\tfrac{3}{2}(\omega_D + \Delta_D' \cos \alpha \cos \delta)  & -\tfrac{\sqrt{3}}{2} \Delta_D' \cos \delta \sin \alpha & 0 & 0 \\
-\tfrac{\sqrt{3}}{2} \Delta_D' \cos \delta \sin \alpha & -\tfrac{1}{2}(\omega_D + \Delta_D' \cos \alpha \cos \delta) & - \Delta_D' \cos \delta \sin \alpha & 0 \\
0	& - \Delta_D' \cos \delta \sin \alpha & +\tfrac{1}{2}(\omega_D + \Delta_D' \cos \alpha \cos \delta) & -\tfrac{\sqrt{3}}{2} \Delta_D' \cos \delta \sin \alpha \\
0	&	0 	& -\tfrac{\sqrt{3}}{2} \Delta_D' \cos \delta \sin \alpha	& +\tfrac{3}{2}(\omega_D + \Delta_D' \cos \alpha \cos \delta) 
\end{pmatrix}.
\end{align}
\end{widetext}
One can diagonalize this matrix;  the eigenvalues correspond to the shifted state energies.  The difference between adjacent $m$-levels, the observed vector-like shift, is
\begin{align}
\Delta_D &= \sqrt{\omega_D^2 + \Delta_D' \cos \delta(2\omega_D \cos \alpha + \Delta_D' \cos \delta)} - \omega_D \\
\intertext{which, for small shifts $\Delta_D' \ll \omega_D$, can be expanded into}
 \Delta_D &=- \cos \alpha \cos \delta \Delta_D' + \frac{\Delta_D'^2}{2\omega_D}(\cos \delta \sin \alpha)^2 + O(\Delta_D'^3). \label{eq:lightshiftAlignError}
\end{align}
We see that the observed energy shift $\Delta_D$ is reduced given finite misalignments and polarization errors.  The first term, linear in $\Delta_D'$ however, is common to both $\Delta_D$ and $\Delta_S$ and therefore cancels in the ratio.  We mechanically constrain misalignment angles to below $5^\circ$, verified by the observed efficiency of optical pumping.  Deviations from true perfect circular polarization are almost always less than 5\%.  For the typical $\Delta_D \approx 10$~kHz light shift on a background magnetic field splitting $\omega_D \approx 2.5$~MHz, the relative error due to Eq.~\ref{eq:lightshiftAlignError} is less than $1 \times 10^{-4}$.  A similar treatment of the tensor shift~\cite{koerber2003thesis} mixed into $\Delta_D$ by a misaligned beam gives a similarly small error.  The relative systematic shift is
\begin{equation}
\frac{\Delta_{D,\text{tensor}}}{\Delta_D} \sim \frac{3}{4} \frac{\Delta_D}{\omega_D} \left(\sin^2 \delta - 2 \sin^2 \alpha \right).
\end{equation}

\subsection{Laser frequency error, spectral impurity, residual fluctuations}
\begin{figure}
\centering
\includegraphics[width=\linewidth]{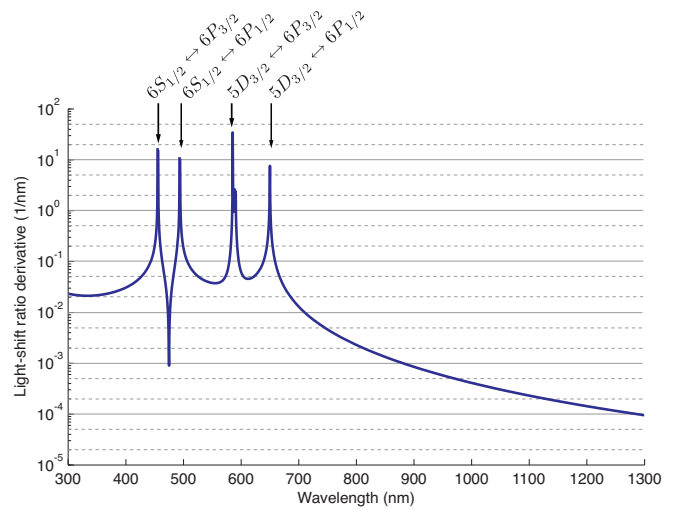}
\caption[An estimate of the light shift ratio \emph{slope}.]{An estimate of the light shift ratio \emph{slope} $|R'|$ with respect to wavelength gives us an idea of how sensitive the measured light shift ratio will be to fluctuations in the laser frequency.  The slope diverges near atomic dipole resonances.  In regions with an expected relative error $|R'| < 0.1$ nm$^{-1}$, a laser with free running stability of $\Delta \lambda = 0.01$~nm contributes a relative systematic error to the light shift ratio of $10^{-3}$.  The region with low $|R'|$ around 470~nm is in-between the $6S_{1/2} \leftrightarrow 6P_{1/2}$ and $6S_{1/2} \leftrightarrow 6P_{3/2}$ transitions.}
\label{fig:LSRslope}
\end{figure}
How accurately must we measure and control the light shift beam wavelength? Consider a quantity $R'$, the relative light shift ratio \emph{slope} with respect to wavelength:
\begin{equation}
R'(\lambda_0) \equiv \left. \frac{1}{R(\lambda_0)} \frac{d \, R(\lambda)}{d \lambda} \right|_{\lambda_0}.
\end{equation}
We expect $|R'|$ to become large only near Ba$^+$ dipole transitions.  Using existing literature values for the barium reduced matrix elements, we plot this function in Figure~\ref{fig:LSRslope}.  Examining it helps us determine the level of laser stability required at particular wavelengths.  If a given laser can be stabilized to within $\Delta \lambda$ over the course of the experiment then the contribution to the expected relative error in the reported light shift ratio is
\begin{equation}
\frac{\sigma_{\lambda}}{R} =  |R'(\lambda_0)| \Delta \lambda.
\end{equation}

The maximum drifts observed using a wavemeter over the months spent collecting data were $\Delta \lambda = $ 0.001~nm and 0.01~nm for the 514.531~nm and 1111.68~nm lasers respectively.  We adjusted the wavelength of the Yb-fiber laser at the beginning of each data run but made no attempt to actively stabilize it.  The Yb-fiber laser is pumped with high power 795~nm diode lasers.  Using a prism we separated out and observed $< 1$~mW of pump light emitting from the fiber.  Any residual pump light that did end up focused on the ion would not likely be well circularly polarized since the polarizing elements we employ are wavelength dependent.  The purity and stability of the 514~nm laser is discussed elsewhere~\cite{sherman2005pml,sherman2007thesis}.  We programmed the data acquisition system to monitor the light shift beam power with a photodiode and halt data taking in the case that the light shift power drifted 5\% above or below a set point.  We employed active laser intensity control during the experiments at 514~nm.

\subsection{Off-resonant, rf ac-Zeeman shifts (i.e.\ from trapping currents)} \label{sec:acZeemanLS}
\begin{figure}
\centering
\includegraphics[width=\linewidth]{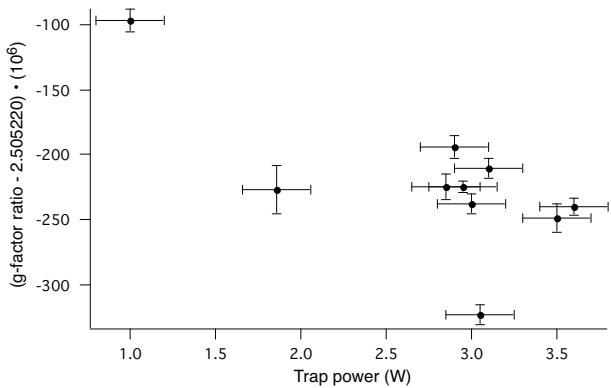}
\caption{Data showing the measured $g$-factor ratio deviation with trap rf strength.  An electrical model shows it is plausible for this shift to be due to the off resonant ac-Zeeman effect.  In~\cite{sherman2007thesis} we argue that the net effect on the measurements of $R$ reported here are at the $5 \times 10^{-4}$ level -- far below the statistical sensitivity.}
\label{fig:gFactorTrapShiftData}
\end{figure}

All our spin resonance measurements are infected with off-resonant \emph{magnetic} light shifts (the ac-Zeeman effect) because measurements of the g-factor ratio
\begin{equation}
G_\text{meas} \equiv \frac{g(6S_{1/2})}{g(5D_{3/2})} = \omega_S / \omega_D
\end{equation}
consistently deviated from the precisely measured value $G = 2.505220(2)$~\cite{knoll1996egf}.  The sign of the deviation changes as we move the resonances above the trap frequency by increasing $B$~\cite{sherman2007thesis}.  Further, as shown in Figure~\ref{fig:gFactorTrapShiftData}, the deviation scales with the trapping rf strength which confirms that rf currents at the trapping frequency $\omega_\text{trap} \approx 9.5$~MHz generate an oscillating magnetic field of sufficient size to shift our single ion resonances.  Such an off-resonant shift can be written in terms of the magnetic dipole interaction strength (a Rabi frequency) $\Omega_\text{trap rf}$ and detuning from resonance.  Including the Bloch-Siegert term we have shifts in the $6S_{1/2}$ and $5D_{3/2}$ spin resonances
\begin{align} \label{eq:trapRfZeemanResShift}
\Delta \omega_S^\text{trap rf} &= -\frac{\Omega_\text{trap rf}^2}{2(\omega_\text{trap} - \omega_S)} + \frac{\Omega_\text{trap rf}^2}{2(\omega_\text{trap} + \omega_S)}, \\
\Delta \omega_D^\text{trap rf} &= -\frac{\Omega_\text{trap rf}^2}{2(\omega_\text{trap} - \omega_D)} + \frac{\Omega_\text{trap rf}^2}{2(\omega_\text{trap} + \omega_D)}.
\end{align}
A plausible model for the source of these shifts are magnetic currents caused by the high voltage trap rf, but not micromotion of the ion in an inhomogeneous magnetic field~\cite{sherman2007thesis}.  We demonstrate that the largest relative effect these shifts have on the light shift ratio $R$ is at the $5 \times 10^{-4}$ level but could be much worse if the trap rf is made nearly degenerate with any spin resonances, or if any light shift was large compared to the Zeeman splittings.

\subsection{On-resonant, rf Bloch-Siegert shift (i.e.\ from spin-flipping field)}
The resonant spin-flip field is itself responsible for an ac-Zeeman light shift through the Bloch-Siegert effect:
\begin{equation}
\Delta \omega_0^{BS} = +\frac{\Omega_\text{spin flip}^2}{4 \omega_0}.
\end{equation}
The effect would cancel in the light shift ratio \emph{if} we probed the light-shifted and unshifted resonances with the same rf power.  In practice we do not since the light-shifted resonances undergo a substantial amount of temporal noise, from fluctuating laser intensity or pointing, that require broadening the resonances with larger spin-flip fields.  $\Omega_\text{spin flip}^D < 2$~kHz and  $\Omega_\text{spin flip}^S < 4$~kHz during our experiments, so we can bound the effect on the resonances to below 1~Hz.  This translates into systematic light shift errors of $\approx 5 \times 10^{-5}$ for the bulk of the data.  Though this effect is well below the statistical sensitivity and dwarfed by other systematics, we attempted to search for any shift that scaled with $\Omega_\text{spin flip}$ by taking special interleaved data runs differing only in spin-flip magnitude by a factor of 25.

\subsection{Correlated magnetic field shifts}
Any change in the magnetic field correlated with the measurement of $6S_{1/2}$ versus $5D_{3/2}$ resonances vanishes entirely in the light-shift ratio.  However, a change in $B$ correlated with the application of the light shift laser does not cancel out.  The only conceivable mechanism is a fluctuation of the magnetic field due to the actuation of the stepper-motor shutter gating the light shift laser.  We try to suppress any such effect with two layers of magnetic shielding, and by placing the shutter more than 1~m from the trap with the solenoid aligned perpendicular to the quantization axis.  A flux-gate magnetometer placed just outside the magnetic shielding showed no change larger than 1~mG correlated with the light shift beam shutter.  Testing for any shift using the ion itself is straightforward:  leaving the light shift laser turned off, we performed a typical light shift ratio data run with narrow $\omega_D$ resonances.  During several dedicated scans, we find that there are no apparent shutter-correlated shifts at the $\pm 2$~Hz level.  The error would be correlated in both $6S_{1/2}$ and $5D_{3/2}$ resonances and larger in the former by the $g$-factor ratio of 2.5.  We put limits on the error in $R$ due to this effect to $\sim 2 \times 10^{-4}$ for 10~kHz type shifts.  Despite being below the statistical sensitivity, we assign this error estimate to each light shift run.

\subsection{Systematic ion displacements}
Any systematic shift in the ion's position due to the application of the light shift laser will lead to a light shift ratio error:  a displaced ion experiences a different light intensity and magnetic field.  Also, the dipole force on an ion in the $6S_{1/2}$ will be in principle different than when in the $5D_{3/2}$ state.  Fortunately the rf trapping force dwarfs these effects by many orders of magnitude.  Previous work \cite{koerber2003thesis} also treats systematic movement of the ion trap loop and fluctuations of the pseudo-potential due to application of the spin-flip rf:  such mechanisms contribute to a systematic shift at a level far below the other effects considered in this section.

\section{The extraction of matrix elements} \label{sec:extractionOfMatrixElements}
As a simple first step, we can turn our results into values for reduced dipole matrix elements, by solving for two desired quantities in Eq.~\ref{eq:lightshiftRatioEstimate}, using literature values of the other required dipole matrix elements as `known' quantities.  Performing this operation using literature values~\cite{gopakumar2002edq} yields
\begin{align*}
| \langle 5D_{3/2} || e r || 6P_{1/2} \rangle | &= 3.14(8) \, ea_0 \\
| \langle 5D_{3/2} || e r || 4F_{5/2} \rangle | &= 4.36(36) \, ea_0,
\end{align*}  
where the errors are due mainly to uncertainties in the literature values. There are additional errors due to terms in Eq.~\ref{eq:lightshiftRatioEstimate} we have not included, such as core excitations and continuum states.  The latter should contribute at the few percent level; studies of the dc-Stark effect in Ba$^+$~\cite{stambulchik1997ehn} that imply the fractional effect of high-$n$ and continuum levels is $\approx 1.4 \times 10^{-4}$ for $6S_{1/2}, m=\pm1/2$ states and $\approx 2.9 \times 10^{-2}$ for $5D_{3/2}, m=\pm 1/2$ states.  Without such corrections, the value we obtain for $| \langle 5D_{3/2} || e r || 6P_{1/2} \rangle |$ is in good agreement with both the theoretical and experimental results summarized in Table~\ref{tab:dipoleMatrixElementTable}.  Our determination of $| \langle 5D_{3/2} || e r || 4F_{5/2} \rangle |$ is the first precision measurement known to us, and is in reasonable agreement with the theoretical estimate.

A more sophisticated approach to analyzing our results, currently underway~\cite{safronova2006pc}, utilizes modern many-body atomic theory at all stages of the calculation.  Such a direct \emph{ab initio} calculation of the light shift ratio $R$ should enable our measurements to provide an exacting test of atomic theory.

\section{Acknowledgments}
This work is supported by NSF grant number PHY-0457320.  The authors thank M.\ S.\ Safronova, B.\ P.\ Das, and B.\ K.\ Sahoo for their discussion of atomic theory issues.

\bibliography{lspaper}
\end{document}